\documentclass[12pt]{article}
\usepackage{natbib}
\usepackage{latexsym}

\begin{document}

\bibliographystyle{apalike}

\baselineskip 18pt

%
%

\newcommand{\ket}[1]{|#1\rangle}
\newcommand{\bra}[1]{\langle#1|}
\newcommand{\opa}[1]{\mbox{\boldmath $#1$}^{\ast}}
\newcommand{\ip}[2]{\langle#1|#2\rangle}
\newcommand{\braket}[2]{\langle#1|#2\rangle}
\newcommand{\op}[1]{\mbox{\boldmath $#1$}}
\newcommand{\lat}{\mathcal{L}}
\newcommand{\alg}[1]{\mathcal{#1}}
\newcommand{\tpv}[2]{\ket{#1}\otimes \ket{#2}}
\newcommand{\tp}[2]{#1\otimes #2}
\newcommand{\ketbra}[2]{|#1\rangle\langle#2|}
\newtheorem{theorem}{Theorem}
\newtheorem{definition}{Defintiion}

%

\oddsidemargin 0in
\evensidemargin 0in
\topmargin 0in

\title{The EPR Experiment:\\A Prelude to Bohr's Reply to EPR}
\author{Michael Dickson\\History and Philosophy of Science\\Indiana University}
\thanks{Thanks to audiences at Indiana University and HOPOS 2000 for 
comments on related talks.  Thanks to Arthur Fine for alerting me to 
some secondary literature.  Thanks to Michael Friedman and Scott 
Tanona for helpful discussions.}
\date{}
\maketitle
\thispagestyle{empty}
\setcounter{page}{1}

\section{Einstein, Podolsky, and Rosen's argument}

\noindent Bohr's \citeyearpar{bohr35a} reply to Einstein, Podolsky,
and Rosen's (EPR's) \citeyearpar{EPR} argument for the incompleteness
of quantum theory is notoriously difficult to unravel.  It is so
difficult, in fact, that over 60 years later, there remains important
work to be done understanding it.  Work by Fine \citeyearpar{fine86a},
Beller and Fine \citeyearpar{beller94a}, and Beller
\citeyearpar{beller99a} goes a long way towards correcting earlier
misunderstandings of Bohr's reply.  This essay is intended as a
contribution to the program of getting to the truth of the matter,
both historically and philosophically.  In a paper of this length, a
full account of Bohr's reply is impossible, and so I shall focus on
one issue where it seems further clarification is required, namely,
Bohr's attempt to illustrate EPR's argument by means of a thought
experiment.  In addition, I shall attempt to clarify a few other
points which, however minor, have apparently contributed to
misunderstandings of Bohr's position.  As the title of this paper
suggests, an account of these few points does not consitute an account
of Bohr's reply, but it is an important step in that direction.

I shall begin by raising several points about EPR's argument, and
especially their example of particles correlated in position and
momentum.  Some of these points have not been sufficiently noticed in
the literature.

Let us begin with a standard, but incorrect, story about EPR's
argument.  Two particles are emitted from a common source, with
momenta $p$ and $-p$, respectively.  For simplicity, we assume that
their masses are the same.  Some time later, particle $1$ encounters a
measuring device, which can measure either its position, or its
momentum.  If we measure its momentum to be $p$, then we can
immediately infer that the momentum of particle $2$ is $-p$.  If we
measure its position to be $x$, then (letting the source be at the
origin) we can immediately infer the position of particle $2$ to be
$-x$.  Now, if we assume that the measurement on particle $1$ in no
way influences the state of particle $2$, then particle $2$ must have
had those properties all along, because it could not obtain them
merely as a result of the measurment on particle $1$.  But quantum
theory cannot represent particle $2$ as having a definite position and
momentum, and therefore quantum theory is incomplete.

EPR do not make this argument.  If they had, Bohr's reply could have
been quite short.  The short reply is to note that in order to make
the requisite predictions, one must know the precise position and
momentum of the source.  Consider, for example, that you have just
measured the momentum of particle $1$ to be $p$.  If you do not know
the momentum of the source, then in particular you do not know in
which frame of reference to apply conservation of momentum.  (Above we
assumed that the source is at rest relative to us, and so we infered
that particle $2$ has momentum $-p$.)  Similarly, consider that you
have just measured the position of particle $2$ to be $x$.  If you do
not know the location of the source, then you cannot say where
particle $2$ is.  It is `the same distance from the source' as
particle $1$, in the other direction, but how far is particle $1$ from
the source?  Unless you know where the source is, you cannot answer
this question.

But if you must know the precise position and momentum of the source
in order to make the inferences, then the uncertainty principle will
always get in the way of EPR's argument.  Suppose, for example, that
you know the precise momentum of the source.  Then you measure the
position of particle $1$.  The EPR criterion for physical reality
says:
\begin{quotation}
    \noindent If, without in any way disturbing a system, we can
    predict with certainty\ldots the value of a physical quantity,
    then there exists an element of physical reality corresponding to
    this physical quantity.  \citep[p.~777]{EPR}
\end{quotation}
But we {\it cannot} predict particle $2$'s position with certainty,
because we do not (and under the circumstances, cannot) know where the
source is.

Good thing, then, that the `standard story' about EPR's argument is
wrong.  We can see immediately that something is wrong with it,
because nowhere did that story mention quantum theory, and yet EPR are
very concerned to present their argument in quantum-theoretic terms
(as they should be).  Indeed, the first part of their paper rehearses
a number of facts about the formalism of quantum theory, presumably so
that they can present their argument in a quantum-theoretic context
(which is what they do).

EPR continue by considering a generic system of two particles and a
pair of generic (but non-commuting) observables on particle $1$, $A$
and $B$.  EPR do not then write down a generic version of the
so-called `EPR state'.  Instead, they merely point out that as a
result of measuring $A$ on particle $1$, particle $2$ may be left in
one state---call it $\psi_{k}(x_{2})$, as they do---while as a result
of measuring $B$ on particle $1$, particle $2$ may be left in quite
another state---call it $\varphi_{r}(x_{2})$, as they do.

At this stage of the argument, EPR might have pointed out that
$\psi_{k}$ and $\varphi_{r}$ are eigenfunctions of {\it some}
observables.  Hence we would be able to predict, with certainty, the
values of two observables as a result of two different measurements
(of $A$ or $B$) on the first system.  One would then have to go on to
show that those observables need not commute.

Instead of continuing with this generic case, however, EPR turn to a
specific example, using the position and momentum observables.  Here
they do add the idea that $\psi_{k}$ and $\varphi_{r}$ can be
eigenfunctions of position and momentum, respectively.  To establish
this claim, they suppose that the total system prior to any
measurements is in the state
\begin{equation}
  \label{eq:eprState}
  \Psi_{\mbox{\scriptsize EPR}}(x_{1},x_{2}) = 
    \int_{-\infty}^{\infty} e^{(2\pi i/h)(x_{1} - x_{2} + x_{0})p} dp,
\end{equation}
where $x_{0}$ is some constant.  EPR then show that
(\ref{eq:eprState}) is a state of perfect (anti-) correlation between
the positions and momenta of the two particles: measuring the momentum
of particle $1$ (hence collapsing the wavefunction for the compound
system!)  leaves particle $2$ in the relevant eigenstate of momentum,
and likewise for position.

So why is the `standard story' inconsistent with this account?  We
have already noted that the `standard story' is not
quantum-mechanical, but the more important point for us here is that
EPR nowhere describe how the compound system is prepared, nor how it
evolves in time.  Indeed, the notion of time never enters their
discussion.  The state $\Psi_{\mbox{\scriptsize EPR}}$---let us call
it the `EPR state'---is a `snapshot' of the compound system at a time. 
Moreover, EPR {\it could not} give us a dynamical description of the
situation, because the EPR state cannot be preserved under Hamiltonian
evolution (unless we introduce an infinite potential, a point that I
will no longer bother to mention).

The reason is familiar, though not usually mentioned in this context. 
The support of $\Psi_{\mbox{\scriptsize EPR}}$ has measure $0$ in
configuration space: $\Psi_{\mbox{\scriptsize EPR}}(x_{1},x_{2})$ is
zero except when $x_{2} - x_{1} = x_{0}$, and so it is a line in the
(two-dimensional) configuration space for the compound system.  Such a
state necessarily spreads under the evolution induced by any
Hamiltonian.  (We are, of course, ignoring the fact that the EPR state
is not in $L^{2}(R^{2})$ in the first place.  Neither EPR nor Bohr
seem to have been concerned about this point.)

Finally, note that there is no Hamiltonian evolution that can take a
generic state $\Phi(x_{1},x_{2})$ to the EPR state (no matter what
$\Phi$ is).  Only a `collapse' of the wavefunction can produce the EPR
state.  Hence we must imagine the EPR state to exist at {\it and only
at} the moment of preparation.

EPR's argument, then, is based on such a state.  They point out that
upon measuring the position of particle $1$, we can predict with
certainty the position of particle $2$, and likewise for momentum.  Of
course, only one of the two measurements can be performed, which
raises the question whether some modal fallacy has been committed.
After all, their argument apparently takes the form:
\begin{enumerate}
  \item Actually: position is measured for particle $1$, and therefore
        (actually) particle $2$ has a definite position.
  \item Possibly: momentum is measured for particle $2$, and therefore
        (possibly) particle $2$ has a definite momentum
  \item Therefore: particle $2$ (possibly? actually?) has a definite
        position and a definite momentm.
\end{enumerate}
In this form, the argument is clearly fallacious (no matter which
modal version of the conclusion you choose).  Of course, the notion of
`non-disturbance' is supposed to help patch up the argument:  although
the circumstances under which we can predict the value of particle
$2$'s position are incompatible with the circumstances under which we
can predict the value of particle $2$'s momentum, the difference
between these circumstances is supposed to make no difference to
particle $2$.

Even with the help of some principle of non-disturbance, it is not
clear, however, that EPR's argument works.  Let us consider, first, a
`weak principle of non-disturbance':
\begin{quotation}
  \noindent {\bf Weak non-disturbance}: if momentum is measured on
  particle $1$ and (therefore, by the criterion for physical reality)
  momentum is definite for particle $2$, then: had we not measured
  momentum on particle $1$, particle $2$ would still have had a
  definite momentum (and likewise, substituting position for
  momentum).
\end{quotation}
This principle is, alas, not enough to get EPR's conclusion.  They
need:
\begin{quotation}
  \noindent {\bf Strong non-disturbance}: if momentum is measured on
  particle $1$ and (therefore, by the criterion for physical reality)
  momentum is definite for particle $2$, then: had we not measured
  momentum on particle $1$ but instead measured its position, then
  particle $2$ would still have had a definite momentum (and likewise,
  switching position and momentum).
\end{quotation}
The weak principle does not entail the strong principle because it
might be impossible (without destroying essential features of the
situation, in particular, our ability to infer properties of particle
2 from the results of measurements on particle 1) both to measure
position on particle $1$ and for momentum to be definite for particle
$2$.  (In terms of the `possible-worlds' semantics for
counterfactuals: while the closest `momentum is not measured'-worlds
to the `momentum is measured and is definite for particle $2$'-worlds
might all be `momentum is definite for particle $2$'-worlds, those
closest worlds may not contain any `position is measured'-worlds, so
that the closest `momentum is not measured but position is'-worlds to
the `momentum is measured and is definite for particle $2$'-worlds
need not be `momentum is definite for particle $2$'-worlds.  Now say
that sentence three times fast.)

Bohr is sometimes understood to deny the strong principle by asserting
that the act of measuring position on particle $1$ `disturbs' in some
strange `semantic' (and non-local) way the very possibility of
particle $2$'s having a definite momentum.  Such a response is
(rightly) taken to be uninteresting philosophically.  In a longer
account of Bohr's reply, I would argue that while Bohr does deny the
strong principle, he does so for more interesting reasons.  Here,
however, I shall only make a few suggestions in that direction.  The
next section contains several observations about EPR's argument and
Bohr's reply.  These remarks are intended to clear the air of some
minor criticisms of Bohr's reply.  In the subsequent section, I shall
discuss Bohr's thought experiment and make some brief suggestions
about how to understand Bohr's reply.

\section{Some Clarifications}

\noindent 1.  {\it EPR speak in terms of a `contradiction'.} Without
calling into question Fine's \citeyearpar{fine86a} logical analysis of
EPR's argument, we may note that they do speak of a `contradiction'
between their criterion of reality and the completeness of standard
quantum theory.  At the end of the first section of their paper,
\cite{EPR} state their conclusion thus: ``We shall show, however, that
this assumption [completeness], together with the criterion of reality
given above, leads to a contradiction''.

As Beller and Fine \citeyearpar{beller94a} argue, Bohr had no problems
with EPR's criterion for physical reality, nor with their account of
completeness, together understood in a fairly conservative sense
(perhaps, in modern terms, as no more than the eigenstate-eigenvalue
link).  Hence the idea that there might be a `contradiction' between
the criterion and completeness would surely have worried Bohr, and
would understandably be the focus of his reply.  No wonder Bohr's
rhetoric focused on `soundness', `rationality', `lack of
contradiction' and `consistency' (cf.  \cite[pp.~3-4]{beller94a}). 
While we may endorse much of what Beller and Fine
\citeyearpar{beller94a} assert to be at the heart of Bohr's general
concerns about consistency, the simple explanation seems to be just
that EPR do, at least at one point, state their conclusion in terms of
a contradiction, a contradiction that was (for reasons that Beller and
Fine explore) threatening to Bohr's own position.

\vskip 18pt \noindent 2.  {\it The EPR argument focuses on the
example.} I mentioned above that EPR begin their discussion in the
abstract and could have finished it there, but they do not, instead
resorting to the example involving position and momentum.  Bohr, too,
focuses on the example.  Indeed, he takes the example to constitute
the argument, writing that ``[b]y means of an interesting example, to
which we shall return below, they [EPR] next proceed to show that
\ldots [the] formalism [of quantum mechanics] is incomplete''
\cite[p.~696]{bohr35a}.  Nobody involved in the debate seems to have
thought that this focus on the example is unwarranted or misleading. 
The point is important for two reasons.

First, it lends greater importance to a proper understanding of Bohr's
attempt to realize the example in a thought experiment.  From a
contemporary standpoint, one might be tempted to suppose that the real
substance of the EPR argument, and of Bohr's reply, is (and was taken by
them to be) in the more abstract discussions (for example, in the early
part of EPR's paper and the mathematical footnote in Bohr's reply).  While
these more abstract discussions can provide important clues to
understanding EPR and Bohr's reply, their mutual focus on the example of
position and momentum suggests that we too focus on that example in order
to understand what is going on.

Second, the focus on the example {\it is}, in the end, unwarranted and
misleading.  Indeed, from a contemporary standpoint, we can see that
EPR chose a particularly unfortunate example to make their point.  As
I shall emphasize again below, the main problem is that position
(unlike momentum) is not a conserved quantity, so that correlations in
position will in general not be maintained under free (or for that
matter, almost any other) evolution.  Bohm's \citeyearpar{bohm51a}
reworking of EPR's argument in terms of a new example (involving
incompatible spin observables) fixes the problem (because spin is
conserved), and it is unclear whether Bohr's reply could work in this
case.  (In any case, his thought experiment is mostly irrelevant to
the Bohmian example.)

\vskip 18pt
\noindent 3. {\it The observables $X_{1}-X_{2}$ and $P_{1}+P_{2}$ can
be determined simultaneously.}  EPR presume that the total momentum 
($P_{1}+P_{2}$) and the distance between the particles ($X_{1}-X_{2}$) 
can be known simultaneously.  There is no obstacle in principle to 
obtaining such knowledge, since the obervables in question are 
compatible (mutually commuting).  Indeed, the EPR state is a 
simultaneous eigenstate of both of these observables.  (Again, we 
ignore the fact that plane waves and delta functions are not, 
strictly speaking, states, i.e., not in $L^{2}(R^{2})$.)

But how might one actually prepare the EPR state, or more generally,
how might one actually determine $X_{1}-X_{2}$ and $P_{1}+P_{2}$
simultaneously?  That is, from a physical point of view, why do these
operators commute?  Note first---and this point is crucial to an
understanding of Bohr's reply---that Bohr insisted that neither
position nor momentum observables have any clear physical meaning
outside of the specification of some frame of reference.  Bohr is
acutely aware of the role that reference frames play in relativity
theory, and believes that their role in the quantum theory is even
more significant---well-specified frames of reference are crucial to
the very meaning of `spatial location' and `momentum'.  Bohr's view
seems to have been that only prior to the discovery of the quantum
theory, and specifically the `essential exchange of momentum' involved
in any interaction, could one dispense with the insistence that
reference frames are essentially involved in the very notion of
`position' and `momentum'.  While a full analysis of Bohr's position
on this point (and most especially of his understanding of the term
`essential exchange of momentum') is out of the question here, it is
worth noting that Bohr insisted upon the necessary role that
well-defined reference frames play in the very definition of the
notion of position.  He writes:
\begin{quotation}
    \noindent Wie von EINSTEIN betont, ist es ja eine f\"ur die ganze
    Relativit\"atstheorie grundlegende Annahme, da{\ss} jede
    Beobachtung schlie{\ss}lich auf ein Zusammentreffen von Gegenstand
    und Me{\ss}k\"orper in demselben Raum-Zeitpunkt beruht und
    insofern von dem Bezugssystem des Beobachters unabh\"angig
    definierbar ist.  Nach det Entdeckung des Wirkungsquantums wissen
    wir aber, da{\ss} das klassische Ideal bei der Bescreibung
    atomarer Vorg\"ange nicht erreicht werden kann.  Insbesondere
    f\"urht jeder Versuch einer raum-zeitlichen Einordnung der
    Individuen einen Bruch der Ursachenkette mit sich, indem er mit
    einem nicht zu vernachl\"assigenden Austausch von Impuls und
    Energie mit den zum Vergleich benutzten Ma{\ss}st\"aben und Uhren
    verbunden ist, dem keine Rechnung getragen werden kann, wenn diese
    Me{\ss}mittel ihren Zweck erf\"ullen
    sollen.\cite[p.~485]{bohr29a}\footnote{In
    \cite[pp.~97--98]{bohr34a}, the passage reads
    \begin{quotation}
	\noindent As Einstein has emphasized, the assumption that any
	observation ultimately depends upon the coincidence in space
	and time of the object and the means of observation and that,
	therefore, any observation is definable independently of the
	reference system of the observer is, indeed, fundamental for
	the whole theory of relativity.  However, since the discovery
	of the quantum of action, we know that the classical ideal
	cannot be attained in the description of atomic phenomena.  In
	particular, any attempt at an ordering in space-time leads to
	a break in the causal chain, since such an attempt is bound up
	with an essential exchange of momentum and energy between the
	individuals and measuring rods and clocks used for
	observation; and just this exchange cannot be taken into
	account if the measuring instruments are to fulfil their
	purpose.
    \end{quotation}
    As Michael Friedman pointed out to me, the translation does not
    perfectly match the original.  For example, rather than ``an
    essential exchange of momentum'' one should probably say ``a
    non-negligible [nicht zu vernachl\"assigenden] exchange''.  These
    subtle differences are important for a full understanding of
    Bohr's view and especially (perhaps) its development, but for our
    purposes here they are not crucial.}
\end{quotation}
Continuing this line of thought, in his reply to EPR
\citeyearpar[p.  699]{bohr35a}, Bohr writes:
\begin{quotation}
    \noindent To measure the position of one of the particles can mean
    nothing else than to establish a correlation between its behavior
    and some instrument rigidly fixed to the support which defines the
    space frame of reference.
\end{quotation}
Bohr is careful to discuss position (and momentum) in these terms, not
speaking of `the position [or momentum]' of a system, but its position
relative to some other system.  For example, at p.~697 of his reply he
speaks not of the uncertainty of the position of a particle, but of
`the uncertainty $\Delta q$ of the position of the particle relative
to the diaphragm'.  The fact that not only position, but also
uncertainty in position, must be discussed relative to a physically
defined reference frame indicates the extent to which, for Bohr, such
reference frames are involved in the very meaning of `position'.

These points are important, because failing to appreciate them fully,
one can be too easily persuaded that passages such as the one above
indicate Bohr's adherence to a rather strong form of operationalism. 
He might, in other words, be suggesting that physical properties are
defined by the operations used to `measure' them.  But given the
history of Bohr's insistence on the role of (physically specified) 
reference frames in quantum theory, we can just as well (and indeed, I
would argue, more fruitfully) read the passage above and others like
it as insisting that a well-defined frame of reference is crucially a
part of the notion of position.

\vskip 18pt
\noindent 4. {\it The observables $X_{1}-X_{2}$, $P_{1}+P_{2}$,
$X_{1}$, and $P_{1}$ are not mutually commuting.}  It is easy to 
suppose that without losing our knowledge of $X_{1}-X_{2}$ and 
$P_{1}+P_{2}$, we may go on to determine either $X_{1}$ or $P_{1}$.  
(This mistake is all the easier if one conceives of the EPR experiment 
in terms of the `standard story' that I outlined above.)  The 
following passage, for example, seems to make this suggesstion:
\begin{quotation}
    \noindent EPR consider a composite system in a state where, at
    least for a moment, both the relative position $X_{1} - X_{2}$ and
    the total momentum $P_{1} + P_{2}$ are co-measurable.  Moreover,
    in EPR both of these quantities are simultaneously determinable
    with either the position or the momentum (not both) of particle 1. 
    \cite[p.~15]{beller94a}
\end{quotation}
However, $X_{1}$ fails to commute with $P_{1}+P_{2}$ and $P_{1}$ fails
to commute with $X_{1}-X_{2}$.  If the EPR situation allowed us to
co-determine both $X_{1}-X_{2}$ and $P_{1}+P_{2}$ with either $X_{1}$
or $P_{1}$, then a great deal more than Bohr's reply would be in
jeopardy.  If we are to determine $X_{1}$, then we must give up our 
knowledge of $P_{1} + P_{2}$, and if we are to determine $P_{1}$, we 
must give up our knowledge of $X_{1} - X_{2}$.

As Beller \citeyearpar[ch.~6]{beller99a} explains, the early Bohr was
very concerned to explain {\it why} it is not possible to observe
simultaneous values for incompatible observables.  I will suggest,
below, that Bohr's reply to EPR continues this discussion, i.e., that
he is, in part, attempting to explain why one cannot measure $X_{1} -
X_{2}$, $P_{1} + P_{2}$, and either of $X_{1}$ or $P_{1}$
simultaneously, within the context of EPR's example.  (Here, then, is
one sense in which Bohr's reply involves themes and argumentative
strategies that he had already used in other cases.)

\section{Bohr's Thought Experiment}

We are now in a position to assess the relevance of Bohr's proposed
thought experiment to EPR's argument.  Bohr's discussion begins with a
rehearsal of two different sorts of experiment.  In the first, there
is a screen with a single slit, ``rigidly fixed to a support which
defines the space frame of reference'' \citeyearpar[p.~697]{bohr35a},
and a particle is fired at the screen.  We assume that the particle's
initial momentum is well-defined.  Bohr asks whether, after preparing
the particle in a state of well-defined position by passing it through
the slit (and thereby, according to de Broglie's relation, rendering
its momentum uncertain), we cannot take into account the exchange of
momentum between the particle and the apparatus, thereby `repairing'
the loss of initial certainty about the momentum.  His answer is `no',
because the exchange of momentum ``pass[es] into this common support''
which, because it {\it defines} the space frame of reference, {\it
must} be taken to be at rest, and so ``we have thus voluntarily [by
fixing the initial screen to the support and taking that support to
define the spatial reference frame] cut ourselves off from any
possibility of taking these reactions separately into account''
(ibid.).  (Recall Bohr's claim that ``just this exchange cannot be
taken into account if the measuring instruments are to fulfill their
purpose'', quoted above.)

If, on the other hand, we allow the initial screen to move freely 
relative to the support, then we can indeed measure the exchange of 
momentum between the particle and the screen, but in so doing, we 
necessarily lose whatever information we might previously have had 
about the location of the initial screen relative to the support, and 
therefore passing the particle through the slit is no longer a 
preparation of its position relative to the support:
\begin{quotation}
    \noindent In fact, even if we knew the position of the diaphragm
    relative to the space frame [i.e., the `support'] before the first
    measurmeent of its momentum, and even though its position after
    the last measurement [required to determine the exchange of
    momentum] can be accurately fixed, we lose, on account of the
    uncontrollable displacement of the diaphragm during each collision
    process with the test bodies, the knowledge of its position when
    the particle passed through the slit. 
    \citeyearpar[p.~698]{bohr35a}
\end{quotation}
Note that two measurements of the momentum of the screen are {\it
required} (in addition to a prior measurement of the momentum of the
incident particle) in order to apply conservation of momentum to the
total system, by which we can determine the momentum of the incident
particle after it has passed through the slit.  Bohr claims that the
second measurement of the momentum of the screen disturbs its position
relative to the support in an `uncontrollable' way, thereby preventing
us from determining its position (relative to the support) at the
moment that the particle passed through the slit.

My aim in making these observations is not to analyze Bohr's claims in
detail.  Such an analysis would include a deeper discussion of Bohr's
notion of a `reference frame', and his notion of `uncontrollable
disturbance', both of which are crucial to a complete understanding of
Bohr's reply.  The aim here, rather, is only to remind the reader of
the broad outlines of Bohr's understanding of the uncertainty
principle, and roughly how he defends that understanding by means of
simple thought experiments.  The main point is that Bohr believes that
the `uncontrollable exchange' of momentum and energy between a
measured system and a measuring apparatus entails that those
experimental situations that allow the determination of a particle's
position relative to a given reference frame forbid the determination
of its (simultaneous) momentum relative to that frame, and similarly,
those experimental situations that allow the determination of a
particle's momentum relative to a given frame---by means of an
application of conservation laws---forbid the determination of its
(simultaneous) position relative to that frame.

Let us turn, then, to Bohr's realization of EPR's particular case.  He
proposes a thought experiment to prepare the EPR state, and to perform
the relevant measurements, as follows:
\begin{quotation}
    \noindent The particular quantum-mechanical state of two free
    particles, for which they [EPR] give an explicit mathematical
    expression, may be reproduced, at least in principle, by a simple
    experimental arrangement, comprising a rigid diaphragm with two
    parallel slits, which are very narrow compared with their
    separation, and through each of which one particle with given
    initial momentum passes independently of the other. 
    \cite[p.~699]{bohr35a}
\end{quotation}
The arrangement as described thus far allows one to prepare the pair
of particles in an eigenstate of $X_{1} - X_{2}$, the eigenvalue
being, of course, the distance between the slits ($x_{0}$ in EPR's
notation).  In order to determine $P_{1} + P_{2}$, Bohr proposes the
following (a continuation of the quotation above):
\begin{quotation}
    \noindent If the momentum of this diaphragm is measured accurately
    before as well as after the passing of the particles, we shall in
    fact know the sum of the components perpendicular to the slits of
    the momenta of the two escaping particles, as well as the
    difference of their initial positional coordinates in the same
    direction.  (ibid.)
\end{quotation}
Thus, at this point in the description of the thought experiment, we
have determined (or prepared) the values of $X_{1} - X_{2}$ and $P_{1}
+ P_{2}$ simultaneously.

The crucial question, now, is how one may go on to measure either 
$X_{1}$ or $P_{1}$, in order to determine either $X_{2}$ or $P_{2}$.  
Concerning the measurement of $X_{1}$, Bohr begins
\begin{quotation}
    \noindent [T]o measure the position of one of the particles can
    mean nothing else than to establish a correlation between its
    behavior and some instrument rigidly fixed to the support which
    defines the space frame of reference.  Under the experimental
    conditions described such a measurement will therefore also
    provide us with the knowledge of the location, otherwise
    completely unknown, of the diaphragm with repect to this space
    frame when the particles passed through the slits.  Indeed, only
    in this way we obtain a basis for conclusions about the initial
    position of the other particle relative to the rest of the
    apparatus.  \cite[p.~700]{bohr35a}
\end{quotation}
Bohr has not yet arrived at his main point, but is here pointing out
that, because the initial screen must be allowed to move freely with
respect to the support (so that conservation of momentum can be
applied to it plus the pair of particles), we do not know where it is
relative to the support until we measure the position of one of the
particles (relative to the support).  After such a measurement, we can
learn the position of the screen, because the particles are located
where the slits in the screen are located.  And once we know where the
screen itself is in relation to the support, we can use our knowledge
of $X_{1}$ to infer the location of the other particle, as Bohr says. 
Note, in particular, that Bohr nowhere supposes that the measurement
of the position of the particle disturbs the screen.

Bohr continues:
\begin{quotation}
    \noindent By allowing an essentially uncontrollable momentum to
    pass from the first particle into the mentioned support, however,
    we have by this procedure cut ourselves off from any future
    possibility of applying the law of conservation of momentum to the
    system consisting of the diaphragm and the two particles.  (ibid.)
\end{quotation}
The consequence, as Bohr notes, is that in fact we {\it lose} the
ability to predict the momentum of the second particle, {\it even if}
we were (counterfactually, of course) to measure the momentum of the
first particle.  In the terms of the first section of this essay, Bohr
has rejected `strong non-disturbance', more or less for the reason
suggested there: a measurement of $X_{1}$ necessarily destroys an
essential feature of the compound system prior to measurement, that
feature being the truth of the conditional: if we were to measure
$P_{1}$, then we could predict (with certainty) $P_{2}$.  (A more
complete analysis of Bohr's position would require a longer discussion
of the logic of counterfactuals, which we cannot pursue here.)

From this point of view, Beller and Fine's \citeyearpar{beller94a}
complaints against Bohr's thought experiment are not quite right. 
They make two complaints.  First, they are unhappy with the fact that,
in Bohr's arrangement, ``we have no choice but to measure $X_{1}$ at
the very moment of passage of the two particles through the first
diaphragm'' \cite[p.~14]{beller94a}.  As I have already pointed out,
however, there is really no choice.  No quantum-mechanical state can
evolve into the EPR state, and the EPR state cannot be preserved by
any time evolution.  Hence it can be the state of a system at, and
only at, the moment of preparation.  We can hardly fault Bohr for this
situation.

Their second complaint arises from the first.  They rightly point out
that Bohr does not describe in any detail how the measurement of
$X_{1}$ is to occur.  Indeed, straightforwrd physical consideration of
the situation seems to imply that any such measurement would involve a
disturbance of the diaphragm with the two slits---either indirectly
(for how could one interact with the particle without `touching' the
diaphragm?)  or directly, by simply fixing the diaphragm to the
support.  Beller and Fine appear to opt for the latter.  After
apparently claiming (as I noted above) that EPR's case allows for the
simultaneous determination of $X_{1} - X_{2}$, $P_{1} + P_{2}$ {\it
and} either $X_{1}$ or $P_{1}$, they write:
\begin{quotation}
    \noindent Bohr's double slit arrangement does not satisfy this
    requirement.  In Bohr's example only one of $X_{1} - X_{2}$ or
    $P_{1} + P_{2}$ could be co-determined together with the variable
    [$X_{1}$ or $P_{1}$] one chooses to measure on particle 1. 
    Indeed, we actually have to change the set-up of the two-slit
    diaphragm depending on whether we intend to measure position or
    momentum on particle 1.  In the first case the two-slit diaphragm
    must be immovable; in the second case it must be moveable. 
    \citeyearpar[p.~15]{beller94a}
\end{quotation}
Mainly because of this situation, Beller and Fine refer to Bohr's
realization of EPR's argument a ``flawed assimilation of EPR to a
double slit experiment'' (ibid., p.~16).

I suggest an alternative account.  According to this account, Bohr
completely ignores the fact---even if it follows from simple physical
considerations---that a measurement of $X_{1}$ implies either a
disturbance of the diaphragm or that it is fixed to the support. 
Instead, he is concerned to point out that a measurement of $X_{1}$
involves an uncontrollable exchange of momentum between particle 1 and
the support that defines the space frame of reference, in {\it
precisely} the same way that it does in the simpler cases discussed
prior to EPR. Hence the momentum of particle 1 becomes undefined, and
hence the total momentum (of the pair of particles) becomes undefined. 
Or to put the point in more Bohrian terms: conservation of momentum
cannot be applied to the compound system, and therefore $P_{1} +
P_{2}$ is undefined, because in order for it to be defined, we must be
able to apply conservation of momentum to the diaphragm plus the two
particles.

At the very least, this account has the merit of following quite
closely Bohr's account of the disturbance.  He does not say that, in
the measurement of $X_{1}$, momentum is exchanged between particle 1
and the diaphragm; nor does he ever suggest, in the EPR arrangement,
that the diaphragm is fixed to the support.  Rather, he says that
``momentum [passes] from the first particle into the mentioned 
support'' \cite[p.~700]{bohr35a}.

Similarly, in his account of what goes wrong when we measure $P_{1}$,
he claims that such a measurement removes the possibility of
determining the location of the diaphragm relative to the support.  He
could have two arguments in mind.  First, along lines suggested by
Beller and Fine, one might argue that any measurement of $P_{1}$ must
involve a disturbance of (exchange of momentum with) the diaphragm,
thereby disturbing its position relative to the support, because the
measurement of $P_{1}$ must occur at the moment of preparation. 
Second, along the lines that are suggested here, one might argue that
since the arrangement {\it requires} the diaphragm to move freely with
respect to the support (lest we be unable to determine $P_{1} +
P_{2}$), the only way to determine the location of the diaphragm
relative to the support would be to measure the position of one of the
particles, relative to the support.  But for reasons that were
discussed prior to the case of EPR, measuring $P_{1}$ `cuts one off'
from the possibility of determining particle 1's (and therefore the
diaphragm's) position relative to the support.

\section{Bohm's version of the argument}

I finish with a brief comment regarding Bohm's \citeyearpar{bohm51a}
alternate realization of the EPR state.  The main point is that Bohm's
realization does not involve position and momentum, but incompatible
spin observables.  There are two essential differences between this
case and Bohr's (and EPR's).  First, spin observables, while in a
sense dependent on the specification of a spatial frame of reference
(because we need to know {\it which} direction is, for example, the
`z' direction), are not bound up as closely with the very notion of a
frame of reference.  In particular, the sort of exchange that must
occur between particle and apparatus in a measurement of spin does not
seem to involve a disturbance of the very reference frame used to
define the notion of `direction of spin'.  Second, spin is a conserved
quantity (unlike position), so that the measurement of spin on one
particle can be made long after the preparation of the particles.

It remains to be seen whether a Bohrian response of the EPR argument
can be worked out in the case of spin.  My suspicion is that the
Bohrian response would at the least require significant revision.  As
far as I am aware, Bohr never reacted, publicly or privately, to
Bohm's proposed thought experiment.  (And, of course, it is more or
less Bohm's version that was eventually performed.)  However, the
investigation of these questions must be preceded by a more complete
account of Bohr's reply to EPR, to which the remarks here are at best
a partial preface.

\end{document}